\documentclass{elsart}
\usepackage{graphics}
\usepackage{psfrag}
\usepackage{epsfig}
\usepackage{epsf}
\usepackage{float}
\usepackage{amssymb,stmaryrd,latexsym}
\usepackage{amscd}
\usepackage{mathrsfs}
\usepackage{times}
\usepackage{color}

\newcommand{\beq}{\begin{equation}}
\newcommand{\eeq}{\end{equation}}
\newcommand{\beqa}{\begin{eqnarray}}
\newcommand{\eeqa}{\end{eqnarray}}
\newcommand{\fet}[1]{\mbox{\boldmath $#1$}}

\newcommand{\ben}{\begin{displaymath}}
\newcommand{\een}{\end{displaymath}}
\newcommand{\be}{\begin{equation}}
\newcommand{\ee}{\end{equation}}
\newcommand{\bea}{\begin{eqnarray}}
\newcommand{\eea}{\end{eqnarray}}

\definecolor{grey}{rgb}{0.3,0.3,0.3}

\usepackage{amsmath}
\usepackage{amssymb}
\usepackage{epsfig}

\begin{document}

\begin{frontmatter}
\title{Weinberg's approach to nucleon-nucleon scattering revisited}

\author{E. Epelbaum$^{1}$, J. Gegelia$^{1,2}$}

{\small $^1$ Institut f\" ur Theoretische Physik II, Fakult\" at f\" ur Physik und Astronomie,\\ Ruhr-Universit\" at Bochum 44780 Bochum, Germany} \\
{\small $^2$ Tbilisi State University, 0186 Tbilisi, Georgia}\\

\begin{abstract}
\noindent 
We propose a new, \emph{renormalizable} approach to nucleon-nucleon scattering in
chiral effective field theory based on the manifestly Lorentz
invariant form of the effective Lagrangian without employing  the
heavy-baryon expansion. For the pion-less case and for the formulation based
on perturbative pions, 
the new approach
reproduces the known results obtained by Kaplan, Savage and Wise. 
Contrary to the standard formulation utilizing the heavy-baryon
expansion, the nonperturbatively resummed one-pion exchange potential can 
be renormalized by absorbing \emph{all} ultraviolet divergences into
the leading S-wave contact interactions. 
We explain in detail the differences to the heavy-baryon formulation
and present numerical results for two-nucleon phase shifts at
leading order in the low-momentum expansion.  
\end{abstract}

\end{frontmatter}

\section{Introduction}
\label{sec1}

The last two decades have witnessed a renewed interest in the 
nuclear force problem and nuclear physics thanks to 
the development and application of effective field theory (EFT) methods. Much of this
research has been influenced by the seminal work of Weinberg \cite{Weinberg:rz} who
was the first to apply chiral perturbation theory (ChPT) to
nucleon-nucleon scattering. 
Using the heavy baryon (HB) formulation of ChPT, he showed that
reducible time-ordered nucleon-nucleon (NN) diagrams  yield enhanced 
contributions to the scattering amplitude as compared to naive 
dimensional analysis. The enhancement can be traced back to 
the appearance of pinch singularities
emerging from the two-nucleon  intermediate states. The enhanced
contributions can be most easily and efficiently resummed by solving the
Lippmann-Schwinger (LS)
equation. The description of low-energy nucleon dynamics, therefore,
naturally reduces to the conventional, quantum mechanical $A$-body
problem where the nuclear forces are defined as a kernel of the
corresponding dynamical equation and can be derived order-by-order in ChPT. 

\begin{figure}
\begin{center}
\epsfig{file=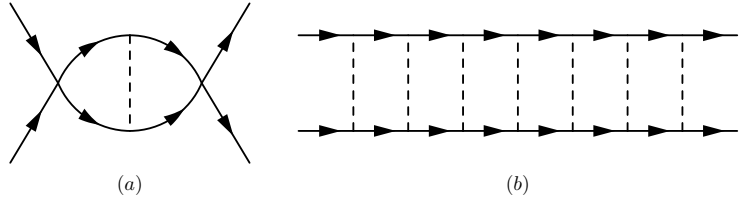,width=0.7\textwidth}
\end{center}
\caption[]{\label{fig:inconsistency} Two examples of iterations of
the lowest-order Lippmann-Schwinger equation.}
\end{figure}

Starting from the pioneering work of Ref.~\cite{Ordonez:1995rz}, this
approach has developed rapidly over the last decades and is nowadays widely
employed in studies of low-energy few- and many-nucleon dynamics and 
nuclear structure calculations, see 
\cite{Epelbaum:2008ga,Machleidt:2011zz,Epelbaum:2012vx} for recent
review articles. 
While offering many attractive features such as simplicity and the
ability to use well-developed machinery to 
treat few- and many-body dynamics, Weinberg's approach suffers 
from being rather intransparent with regard to renormalization. 
One issue is related to the fact that iterations of
the truncated NN potential within the LS equation generate
contributions to the amplitude beyond the order one is working. 
These higher-order terms generally involve ultraviolet (UV)
divergencies which cannot be absorbed by counter terms (contact
interactions) included in the truncated potential so that one needs to
employ a finite UV cutoff $\Lambda$ of the order of a natural hard
scale, say $\Lambda \sim \Lambda_\chi \sim M_\rho$ \cite{Lepage:1997cs}. 
While subleading and higher-order corrections to the potential do
not have to be resummed in Weinberg's power
counting scheme and can be treated perturbatively, the LS equation for
the leading-order (LO) potential already turns out to be not
renormalizable (in the usual sense). 
In particular, infinitely many counter terms are needed to absorb 
UV divergences emerging from iterations of the LO long-range
potential due to one-pion exchange (OPE), whose singular
$1/r^3$-piece generates UV divergencies in all spin-triplet partial waves. 
This problem, in fact,  shows up in every spin-triplet partial wave.  
To be specific, consider the lowest-order potential in Weinberg's approach,  
\beq
\label{LO}
V^{\rm LO} = - \frac{g_A^2}{4 F_\pi^2}  \fet \tau_1 \cdot
\fet \tau_2 \frac{\vec \sigma_1 \cdot \vec q \; \vec \sigma_2 \cdot
  \vec q}{\vec q \, ^2 + M_\pi^2} + C_S + C_T \vec \sigma_1 \cdot \vec \sigma_2\,,
\eeq
where $g_A$, $F_\pi$ and $M_\pi$ are the nucleon axial-vector coupling,
pion decay constant and the pion mass, respectively,  $\vec
\sigma_i$ ($\fet \tau_i$) denote the spin (isospin) Pauli matrices of the
$i$-th nucleon and $\vec q = \vec p \,  ' - \vec p$ is the nucleon momentum
transfer. It is easy to see by dimensional arguments that the $2n$-th
iteration of this potential in the LS equation generates a logarithmic divergence
$\propto (Q m)^{2n}$ \cite{Savage:1998vh}, where $m$ is the nucleon
mass and $Q$ denotes the generic soft scale
corresponding to external three-momenta of the nucleons and $M_\pi$.  
In the $^1S_0$ channel, where the singular tensor
part of the OPE potential vanishes, the coefficients in
front of the logarithmic divergences do not involve external momenta   
and can be absorbed by derivative-less contact operators with
multiple insertions of $M_\pi^2$. The potentially enhanced
contributions of these higher-order $M_\pi^2$-dependent operators 
might become an issue if one is interested in the quark mass
dependence of nucleon-nucleon scattering but do not affect the predictive power
of the theory in terms of describing the energy dependence of the phase
shift at the physical values of the quark masses. 
On the other hand, in spin-triplet channels,  the coefficients of
the logarithmic divergences do involve powers of external
momenta.\footnote{If cutoff regularization is employed, which is
  normally the case in calculations with non-perturbative pions,
  one also needs to keep track of power-like
  divergences. The coefficients in front of power-like
  divergences generally also involve powers of external momenta.}
Their removal requires the inclusion of an infinite number of
higher-derivative contact interactions. For example, calculating the diagrams
of Fig.~\ref{fig:inconsistency} in dimensional regularization with
$n$ spatial dimensions 
one finds the divergent parts proportional to $ m^2 M_\pi^2/(n-3)$
for graph (a) and, among many other divergent terms, $m^6 (\vec p \, ^6 + \vec
p\,  '  {}^6 )/(n-3)$ for graph (b) \cite{Savage:1998vh}.   
The appearance of such divergences seems to indicate that the same 
enhancement, which is responsible for non-perturbativeness of the OPE
potential\footnote{One way to justify the need to resum the OPE 
  and the LO contact operators is to treat the nucleon mass as a separate scale in the problem,
  which is assumed to be much larger than
  the chiral symmetry breaking scale \cite{Weinberg:rz}.}, also applies to higher-order short-range operators. 
This feature is sometimes referred to as inconsistency of Weinberg's approach.
It should, however, be understood that this 
issue does not
affect the predictive power of Weinberg's approach with regard to
describing the energy dependence of the scattering amplitude provided
a suitably chosen finite cutoff is employed along the lines of
Ref.~\cite{Lepage:1997cs}. The 
predictive power of such a framework can
be understood in terms of the modified effective range expansion
\cite{Epelbaum:2009zz,MERE} and
relies on the knowledge of the
long-range tale of the interaction driven by the pion exchange.  
We also emphasize that there is no
consensus on the relevance of the inconsistency issue for renormalized
contributions to the scattering amplitude, see e.g.~Ref.~\cite{Gegelia:2004pz}. 

The possibility to remove the UV cutoff  $\Lambda$ from the LS
equation by enforcing the limit $\Lambda \to \infty$ (or,
equivalently,  $\Lambda \gg \Lambda_\chi$) non-perturbatively was also
explored by several authors, see
e.g.~\cite{PavonValderrama:2005gu}. It is possible to obtain a
finite, manifestly non-perturbative solution of the LS equation with
a singular $1/r^3$-potential by including one/no contact operator in
each attractive/repulsive channel \cite{PavonValderrama:2005gu}. 
Unless all UV divergences emerging from iterations of the LS equation
are absorbed by counter terms, such a procedure is expected to violate 
the low-energy theorems and is incompatible with the principles of EFT
\cite{Epelbaum:2009sd}. 
Presently, there is still no consensus in the community on the 
most consistent and efficient way to organize the chiral expansion in
the few-nucleon sector, see
\cite{Lepage:1997cs,PavonValderrama:2005gu,Epelbaum:2009sd,Nogga:2005hy,Epelbaum:2006pt,Long:2007vp,Yang:2009pn,Birse:2010fj,Valderrama:2009ei,Valderrama:2011mv,Long:2011xw} 
for samples of different views and formulations. 

Treating the exchange of pions perturbatively as suggested
by Kaplan, Savage and Wise (KSW) \cite{Kaplan:1998tg} obviously allows one
to avoid the above-mentioned inconsistency and, at the same time, provides a
transparent and analytically solvable EFT framework for 
nucleon-nucleon scattering.   
In the KSW approach, the OPE potential
is shifted to next-to-leading order (NLO). The LO NN amplitude, therefore, becomes
renormalizable perturbatively as well as non-perturbatively. 
Corrections beyond LO are treated 
as perturbations. While the KSW scheme is free of any inconsistencies
with respect to renormalization, it  was shown not to converge in low
spin-triplet partial waves
\cite{Gegelia:1998ee,Cohen:1998jr,Fleming:1999ee}, see however Ref.~\cite{Beane:2008bt}
for a modified formulation. 

In this paper we identify the origin of
non-renor\-mali\-za\-bi\-li\-ty of the LO NN
amplitude in Weinberg's approach with the non-relativistic
expansion of the nucleon propagators. 
It is by now well established in the single-baryon sector of ChPT that 
the chiral power counting can be maintained without relying on the
non-relativistic or HB expansion, see Ref.~\cite{Bernard:2007zu} and references
therein. We propose here a 
formulation of chiral EFT for NN scattering with non-perturbative
pions based on the manifestly Lorentz-invariant form of the
effective Lagrangian which is
consistent and renormalizable. We will refer to this
formulation as the modified Weinberg approach.  
We demonstrate in the case of perturbative pions  
how renormalization is carried out in this framework and 
present cutoff-independent results for NN phase shifts
at LO in the modified Weinberg approach. 
Our paper is organized as follows: In section \ref{sec2} we describe
the framework and specify the dynamical
equation for the NN amplitude. Based on the (modified) KSW approach at
NLO in the EFT expansion, we demonstrate in
section \ref{sec3} the consistency of this scheme with respect to the
power counting. The results for non-perturbative pions 
at LO in the modified Weinberg approach  are presented in section
\ref{sec4}. Finally, our findings are summarized in section
\ref{sec5}.

\section{The framework}
\label{sec2} 

To derive the dynamical equation for NN scattering in chiral EFT we  
follow closely the procedure of Ref.~\cite{Djukanovic:2006mc} 
(but refrain from using high-derivative regularization and performing
expansion in inverse powers
of the nucleon mass $m$). We start with the manifestly Lorentz-invariant effective Lagrangian for pions and nucleons. It is
organized in a derivative and quark-mass expansion and consists of the
purely mesonic, pion-nucleon and nucleon-nucleon parts,
whose lowest-order contributions 
can be found e.g.~in
Refs.~\cite{Djukanovic:2006mc,Gasser:1984yg,Gasser:1988rb}, see also
\cite{Bernard:2007zu} and references therein.
Following Weinberg \cite{Weinberg:1966jm}, we employ
time-ordered perturbation theory without performing non-relativistic 
expansion to calculate NN scattering amplitude. 
We decompose the numerator of the standard fermion propagator  as
\beq
p\hspace{-.45em}/\hspace{.1em}+m
= 2\,m\,P_+   + \left(p
\hspace{-.45em}/\hspace{.1em}-m\,v \hspace{-.45em}/\hspace{.1em}\right),\label{Sfpexpanded}
\eeq
where $P_+ \equiv  (1+v \hspace{-.45em}/\hspace{.1em})/2$ with
$v=(1,0,0,0)$, and identify the second terms as a higher-order
correction (to be included perturbatively). For NN scattering the 
two-nucleon intermediate
states generate enhanced contributions \cite{Weinberg:rz}.
Therefore, defining the two-nucleon-irreducible diagrams as the effective
potential $V$, the NN off-mass-shell scattering amplitude $T$ satisfies the integral equation
written symbolically as
\begin{equation}
T = V + V \,G \ T \,,\label{EqForGF}
\end{equation}
where $G$ is the two-nucleon propagator.
Substituting the expansions of   $V$, $G$ and   $T$ in a small parameter (pion mass or small momenta)
in Eq.~(\ref{EqForGF}), the leading-order contribution emerges from
solving the equation 
non-perturbatively
\begin{equation}
T_0 = V_0 + V_0\, G_0\, T_0\,, \label{LOgeq}
\end{equation}
while corrections are calculated perturbatively using the solution to the LO equation.

The physical amplitude is obtained from the off-shell amplitude $T$ 
via $Z_\psi^2\, \bar u_3\bar u_4\,T\,u_1 u_2$
where $Z_\psi$ is the residue of the nucleon propagator and $u_i$, $\bar u_i$ are Dirac spinors.
To determine the physical amplitude order-by-order we expand the Dirac
spinors in small quantities as
\beq
u =  u_0 + u_1 + u_2 +\cdots\,, \quad \quad
\\
\bar u  =  \bar u_0 +\bar u_1 +\bar u_2 +\cdots\,,
\label{deqsolexpanded}
\eeq
where
\begin{equation}
u_0(p) = P_+\, u(p)\,, \ \ \ \bar u_0(p)  =  \bar u(p)\,P_+\,, \ \cdots.
\label{spellout}
\end{equation}
Consequently, the lowest-order on-shell amplitude requires the
knowledge of the quantity $\tilde T_0 = P_+ P_+\, T_0 \,P_+ P_+$
which fulfills the integral equation 
\begin{equation}
\tilde T_0= \tilde V_0+\tilde V_0\, G\,\tilde T_0 \,,
\label{LOgeqPTilde}
\end{equation}
where $\tilde V_0=P_+ P_+\, V_0 \,P_+ P_+$ is the projected potential.
In Weinberg's approach, the projected lowest-order potential consists
of derivative-less contact interactions and the OPE piece. Following
Ref.~\cite{Djukanovic:2006mc},  we choose to treat corrections to the static OPE
potential perturbatively so that the LO potential takes the form of
Eq.~(\ref{LO}) after switching to the two-component Pauli spinors. 
In the center-of-mass frame with incoming (outgoing)  three-momentum of
the nucleons $\vec p$ ($\vec p\, '$), 
the LO equation  takes the form 
\beq
\tilde T_0\left(
\vec p\,',\vec p \, \right)=\tilde V _0 \left(
\vec p\,',\vec p \, \right) - \frac{m^2}{2}\int \frac{d^3 \vec k}{(2\,\pi)^3} 
 \frac{\tilde V _0 \left(
\vec p\,',\vec k \, \right) \ \tilde T _0 \left(
\vec k,\vec p\right)}{\left(\vec k^2+m^2\right)\left(
p_0-\sqrt{\vec k^2+m^2}+i \epsilon\right)}
\,, 
\label{MeqLOk0integrated}
\eeq
where we have suppressed the spin and isospin indices. 
For the half-off-shell kinematics, $p_0$ is given by  $p_0 = \sqrt{\vec p \, ^2 + m^2}$.  
This equation is, of course, not new and was proposed for the
first time in Ref.~\cite{Kadyshevsky:1967rs}. This and many other equations of a similar
type, all 
satisfying relativistic elastic unitarity,
have been extensively studied in the literature in the context of
three-dimensional reductions of the Bethe-Salpeter equation, see
e.g.~Ref.~\cite{Woloshyn:1974wm}.

It is obvious that Eq.~(\ref{MeqLOk0integrated})
has a milder UV behavior than the LS equation, which emerges from
keeping the leading term in the $1/m$-expansion of the integrand. 
It is perturbatively
renormalizable in the sense that all UV divergences generated by its
iterations are absorbable into the low-energy constants (LECs) $C_{S,T}$ in Eq.~(\ref{LO}).
Nonperturbatively, the UV behavior of this equation will be addressed
in section \ref{sec4}.  While all UV divergences emerging from iterations of
Eq.~(\ref{MeqLOk0integrated}) can be removed by the $C_{S,T}$-counter
terms, the resulting $\Lambda$-independent expression for the
amplitude still violates the power counting due to the explicit appearance of
the  nucleon mass in the integrand  and requires additional, finite
renormalization. This can be achieved by subtractions
which remove the positive powers of $m$ and restore the chiral
power counting, see Ref.~\cite{Fuchs:2003qc} for an extensive discussion. 
Such additional, finite subtractions in the case
at hand only affect the lowest-order contact interactions 
of the $C_{S,T}$ type. An explicit example of renormalization 
will be given in the next section for the case of perturbative pions.

At low energies, the two-nucleon scattering amplitude calculated 
based on the manifestly Lorentz-invariant effective Lagrangian 
can, of course, be expanded in
inverse powers of the nucleon mass. In the HB approach, 
the $1/m$-expansion is carried out already at the level of the Lagrangian
which implies that one first expands the integrand of Eq.~(\ref{MeqLOk0integrated})
and then calculates the scattering amplitude. This is legitimate in
perturbative calculations (apart from some well-known cases such as
e.g.~the scalar form factor of the nucleon, in which the HB approach
fails to reproduce the analytic structure of the relativistic
loop diagrams for soft momenta). We will show in
the next section that the two ways of doing $1/m$-expansion lead to
the same result for the case of perturbative pions in the KSW
framework. 
However, care is required when resumming the pion exchange potential 
in the Weinberg approach. The leading nonrelativistic
approximation of the two-nucleon propagator 
\beq
\frac{m^2}{2 \left(\vec k^2+m^2\right)\left(
p_0-\sqrt{\vec k^2+m^2}+i \epsilon\right)}
\stackrel{\rm }{\longrightarrow}
\frac{m}{\vec p \, ^2 - \vec k^2 + i \epsilon}
\eeq
obviously has a more singular UV behavior compared to the one
appearing in Eq.~(\ref{MeqLOk0integrated}).  
As a consequence, in contrast to the approach presented above and corresponding to
Eq.~(\ref{MeqLOk0integrated}), an infinite set of
counter terms needs to be included in the nonrelativistic framework if one aims
at removing the UV cutoff by taking the limit $\Lambda \to \infty$.

\section{Perturbative pions at next-to-leading order}
\label{sec3} 

\begin{figure}
\epsfig{file=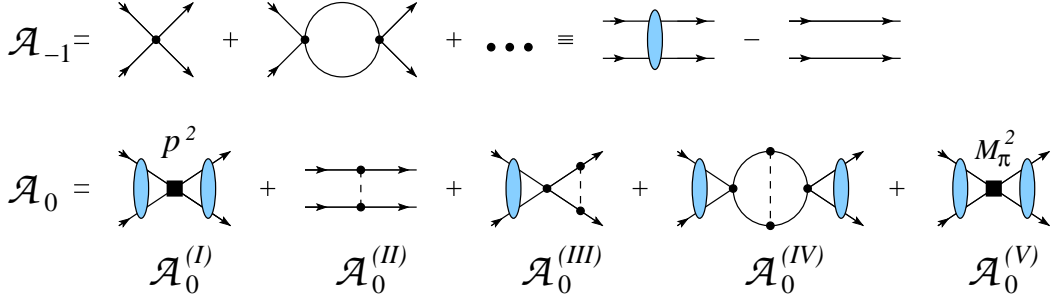,width=\textwidth}
\caption[]{\label{fig:ksw} 
The leading and subleading contributions to
the NN scattering amplitude in the KSW approach. The solid dots denote
the lowest-order contact operators and the leading pion-nucleon
vertex $\propto g_A$ while the filled squares refer to the subleading
contact terms proportional to $p^2$ or $M_\pi^2$. }
\end{figure}

It is instructive to apply Eq.~(\ref{MeqLOk0integrated}) 
to the case of perturbative pions where all
calculations can be carried out analytically. In order to facilitate
the comparison with the original work by Kaplan, Savage and Wise 
based on the HB effective Lagrangian, we adopt here the 
normalization of the amplitude of Ref.~\cite{Kaplan:1998tg}.  
The expansion for the S-wave scattering amplitude in the KSW approach 
has the form 
\begin{equation}
{\cal A}={\cal A}_{-1}+{\cal A}_{0}+{\cal A}_{1}+\cdots \,,
\label{Aexpansion}
\end{equation}
where the subscript indicates the power of the soft scale  $Q$.  The leading-order 
contribution ${\cal A}_{-1}$ emerges from resummation of the LO
contact interactions as shown in Fig.~\ref{fig:ksw}. 
Using the two-nucleon Green function from Eq.~(\ref{MeqLOk0integrated}), the LO
amplitude  has the form
\begin{equation}
{\cal A}_{-1}  = \frac{-C_0}{1-C_0\,I(p)}\,,
\label{Aminus1}
\end{equation}
where the dimensionally regularized (DR) integral $I(p)$ is given  in $n$ spatial dimensions by 
\begin{eqnarray}
I(p) & = & \frac{m^2}{2}\,\frac{\mu^{3-n}}{(2\,\pi)^n}\,\int \frac{d^n
k}{\left[k^2+m^2\right]\left[p_0-\sqrt{k^2+m^2}+i\,0^+\right]}\nonumber\\
& = & \frac{1}{8 \pi^2}
\bigg[ - \bigg( \bar  \lambda + 2  - 2  \ln
   \frac{m}{\mu } \bigg)
m^2 
\nonumber\\ &&{} 
- \frac{m^2}{ \sqrt{m^2+p^2}} \bigg(\pi m +2 i \pi
   p 
- 2 p
    \sinh ^{-1}\left(\frac{p}{m}\right) \bigg) \bigg] + \mathcal{O}(n-3)
\,,
\label{Ieucldef}
\end{eqnarray}
with the divergent quantity $\bar \lambda$ defined as $\bar\lambda
\equiv  -1/(n-3)-\gamma -\ln(4 \pi )$ and $\mu$ being the scale
parameter of DR.  Further, the (bare) LEC $C_0$ is 
simply the properly normalized linear combination of $C_{S,T}$.
Here and in what follows, we use the notation $p \equiv | \vec p \,
|$,   $k \equiv | \vec k \,|$.
Renormalization of ${\cal A}_{-1}$ is achieved by subtracting the loop
integral at $p^2 = - \nu^2$ with $\nu$ chosen to be of order $\mathcal{O} (Q)$, 
\beq
I_{\rm R}(p,\nu) = I(p)-I(i\,\nu)
=-\frac{m (\nu +i\,p)}{4 \pi }  +{\cal O}(p^2,\nu^2)\,,
\eeq
and replacing $C_0$ by $C_0^{\rm R} (\nu )$ which yields 
\begin{equation}
{\cal A}_{-1}  = \frac{-C_0^{\rm R} (\nu )}{1-C_0^{\rm R} (\nu)\,
  I_{\rm R}(p,\nu) } \,.
\label{Aminus1}
\end{equation}
Notice that while just using DR in combination with $\overline{\mbox{MS}}$ would be
sufficient to render the expressions finite, one additional finite
subtraction would have to be performed in order to remove from $I_{\rm R}^{\small
  \overline{\mbox{MS}}}$ terms of
order $\sim m^2$ (i.e. $\mathcal{O} (Q^0)$) which violate the power
counting. The renormalized expression for ${\cal A}_{-1}$ clearly
agrees with the KSW result of Ref.~\cite{Kaplan:1998tg} modulo 
higher-order terms emerging from the $1/m$-expansion of $I_{\rm R}
(p^2,  \nu^2)$.   

The first corrections to the scattering amplitude are generated by diagrams
shown in the second line of Fig.~\ref{fig:ksw}.  The renormalized
contributions of the dressed subleading contact operators 
have the form 
\beqa 
{\cal A}_{0}^{(I)} &=& {\cal A}_{-1}^2
\left[\frac{C_{2}^{\rm R}\,m^2 \left(2 m^2+p^2-2 m
   \sqrt{m^2+p^2}\right) }{8 \pi
   \,C_0^{\rm R}}-\frac{2\,C_{2}^{\rm R}p^2}{(C_0^{\rm
     R})^2}\right],  \nonumber \\ 
{\cal A}_{0}^{(V)} &=& -\frac{D_2^{\rm R}  M_\pi^2}{(C_0^{\rm R})^2}\, {\cal
A}_{-1}^2 \,, 
\eeqa
where $C_{2}^{\rm R} \equiv C_{2}^{\rm R} (\nu )$, $D_{2}^{\rm R} \equiv D_{2}^{\rm R} (\nu )$ denote the corresponding
renormalized LECs and the two subtraction points in ${\cal A}_{0}^{(I)}$ are set to zero
for the sake of convenience. 
The details of the calculation will be given in a
separate publication. 
While ${\cal A}_{0}^{(V)}$ agrees with
the HB result of Ref.~\cite{Kaplan:1998tg} (modulo higher-order corrections from
$I_{\rm R} (p,\nu) $),  the HB result for ${\cal A}_{0}^{(I)}$
is entirely given by the second term in square brackets.  
Given the scaling of the renormalized LECs
$C_0^{\rm R} \sim \mathcal{O} (Q^{-1})$ and  $C_2^{\rm R} \sim
\mathcal{O} (Q^{-2})$ \cite{Kaplan:1998tg}, one observes that the first term in the
brackets is of order $\sim Q^3$ while the second one is of order $\sim
Q^2$. Both approaches, therefore,  again lead to the same
result modulo corrections of a higher order.  

We now discuss the contributions involving pions. The second diagram
in Fig.~\ref{fig:ksw} simply yields the S-wave projected OPE
potential, 
\beq
{\cal A}_{0}^{(II)} =  \frac{g_A^2}{4
F_\pi^2}\,\left(-1+\frac{M_\pi^2}{4\,p^2}\ln\frac{M_\pi^2+4
p^2}{M_\pi^2}\right). 
\eeq
Notice that at the order we are working, there is no need to
distinguish between the physical and the chiral-limit values of the
LECs such as $g_A$, $m$ and $F_\pi$.   
The renormalized contribution of the third diagram reads 
\beqa
{\cal A}_{0}^{(III)} &=& \frac{g_A^2}{2 F_\pi^2}\,{\cal A}_{-1}
\left[I_{\rm R} (p,\nu)
-M_\pi^2 I_{\rm 1 loop} (p)\right],\nonumber\\
I_{\rm 1 loop} (p)&=& \frac{m^2}{2}\int
\frac{d^nk}{(2\,\pi)^{n}}\frac{1}{\left[k^2+m^2\right]\left[p_0-\sqrt{k^2+m^2}+i\epsilon\right]
\left[(k-p)^2+M_\pi^2\right]}\nonumber\\
&=& -\frac{m}{8 \pi  p} \left[\tan ^{-1}\left(\frac{2 p}{M_\pi}\right)+\frac{i}{2}
\ln \frac{M_\pi^2+4
   p^2}{M_\pi^2}\right]+{\cal O}\left(\frac{p}{m},\, \frac{M_\pi}{m}\right)\,.
\eeqa
Again, this agrees with the HB KSW result modulo terms of a higher order
emerging from $1/m$-expansion of $I_{\rm R} (p,\nu)$ and $I_{\rm 1
  loop} (p )$. Finally and most interestingly, for the fourth diagram 
we obtain
\beq
{\cal A}_{0}^{(IV)} = \frac{g_A^2}{4 F_\pi^2}\,{\cal
A}_{-1}^2 \left[ M_\pi^2 I_{\rm 2 loop}-I_R(p,\nu)^2\right],
\eeq
where the scalar two-loop integral has the form
\beqa
\label{twoloop}
I_{\rm 2 loop} &=& \frac{m^4}{4}\int
\frac{d^nk_1d^nk_2}{(2\,\pi)^{2\,n}}\frac{1}{\left[k_1^2+m^2\right]\left[p_0-\sqrt{k_1^2+m^2}+i\epsilon\right]
\left[k_2^2+m^2\right]
}\nonumber\\
&&{} \times
\frac{1}{\left[p_0-\sqrt{k_2^2+m^2}+i\epsilon\right]\left[(k_1-k_2)^2+M_\pi^2\right]}\\
&=& \frac{m^2}{16 \pi ^2} \left[\frac{\ln 8}{4}-\frac{2 G}{\pi }-\frac{7 \zeta
(3)}{2 \pi
   ^2}-\frac{1}{2} \ln
   \frac{M_\pi^2+4 p^2}{m^2}+i \tan
   ^{-1}\left(\frac{2 p}{M_\pi}\right)\right] + \ldots \nonumber 
\eeqa
where $G\approx 0.916$ is Catalan's constant and the ellipses refer to terms of
order $pm$, $M_\pi m$ and
higher. The HB result of Ref.~\cite{Kaplan:1998tg} for this diagram in our
notation has the following form
\beq
\label{twoloopHB}
{\cal A}_{0, \; \rm HB}^{(IV)} = \frac{g_A^2 m^2}{64 \pi^2  F_\pi^2}\,
{\cal A}_{-1}^2 \left(M_\pi^2 \left[-\frac{1}{2} \ln
   \frac{M_\pi^2+4 p^2}{\nu^2} + i \tan
   ^{-1}\left(\frac{2 p}{M_\pi} \right)+1 \right] - (\nu + i p)^2 \right).
\eeq 
Clearly, the difference in the constant terms in the square brackets of
Eqs.~(\ref{twoloop}) and (\ref{twoloopHB}) can be compensated by a
finite shift of the LEC $D_2^{\rm R}$. While all relevant terms
non-polynomial in $M_\pi^2$ and $p^2$ are exactly the same in both cases,
the HB result features a logarithmic dependence of the renormalization
scale which reflects the overall logarithmic divergence of the
integral $I_{\rm 2 loop}$ when the integrand is approximated by the
leading term in the $1/m$-expansion. It is, therefore, \emph{necessary} to include
the $D_2 M_\pi^2$ counter term in the HB approach at the same level
as the diagram (a) of Fig.~\ref{fig:inconsistency}, which appears at LO in the Weinberg
approach, in order to absorb the corresponding logarithmic
divergence. This is, in fact, the first manifestation of the above-mentioned 
inconsistency issue of the Weinberg approach. 
In contrast, the original integral $I_{\rm
  2 loop}$ is finite and fulfills the power counting without any
additional subtractions. Consequently, from the renormalization point of view,
there is no need to promote the $D_2 M_\pi^2$-term to LO if the OPE
potential  is treated
nonperturbatively at the same footing as the $C_0$-term within the
modified Weinberg approach proposed in this paper.  The same
arguments also apply to more complicated diagrams with higher
iterations of the OPE potential.

\section{Non-perturbative pions at leading order}
\label{sec4} 

We now turn to the case of nonperturbative pions and numerically solve
Eq.~(\ref{MeqLOk0integrated}) in the partial wave basis for the LO
potential given in Eq.~(\ref{LO}). We employ a mo\-men\-tum-space
cutoff $\Lambda$ when integrating over $k$ in order to regularize the
divergent integrals.  As discussed in the
previous sections, the LO equation  (\ref{MeqLOk0integrated}) is
perturbatively renormalizable so that one can safely remove the cutoff 
by taking the limit $\Lambda \to \infty$ in any iteration. 
Nonperturbatively, the UV behavior in  Eq.~(\ref{MeqLOk0integrated})
can be understood by approximating the two-nucleon propagator for $k\to\infty$ via
\beqa
\label{propagatorUV}
\frac{1}{\left(k^2+m^2\right)\, \left(
p_0-\sqrt{k^2+m^2}+i\,\epsilon\right)} 
&=& \frac{p_0+\sqrt{k^2+m^2}}{\left(k^2+m^2\right)\, \left(
p^2 - k^2 + i \epsilon\right)} \nonumber \\ 
&\rightarrow& \frac{1}{k\, \left(
p^2-k^2+i\,\epsilon\right)} \,.
\eeqa
The UV behavior of this equation in the partial-wave decomposed form coincides 
with the one of the LS equation in $2+1$ space-time dimensions. 
The OPE part of the potential therefore behaves in coordinate space for $r\to 0$  as
$\sim 1/r^2$  in $2+1$ space-time dimensions. It is well known that the LS 
equation does not possess a unique solution if the strength of the
attractive $1/r^2$ potential exceeds some critical value which depends
on the partial wave, see \cite{Frank:1971xx} for more details. The same sort of non-uniqueness emerges 
in the context of the
Skornyakov--Ter-Martirosyan equation \cite{skornyakov,danilov} which has
also been addressed from the EFT point of view \cite{Bedaque:1998kg}. 
In the case at hand, we found that the non-unique solutions only
appear in the $^3P_0$ partial wave. 
\begin{figure}
\begin{center}
\epsfig{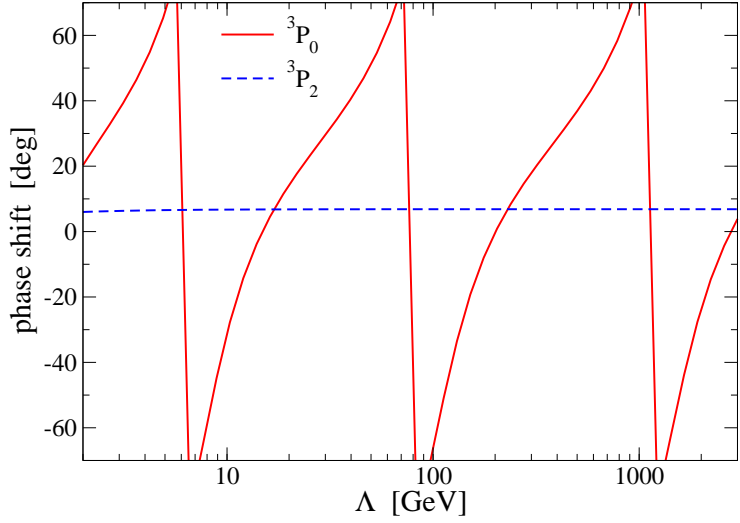}
\end{center}
\caption[]{\label{fig:limcyc} $^3P_0$ and $^3P_2$ phase shifts at
  $E_{\rm lab} =100$ MeV versus the cutoff $\Lambda$ obtained from 
  equation (\ref{MeqLOk0integrated}) with the OPE
  potential.  }
\end{figure}  
This situation is visualized in Fig.~\ref{fig:limcyc} where we compare the dependence of
the $^3P_0$ and $^3P_2$  phase shifts on the UV cutoff $\Lambda$ at the fixed energy of $E_{\rm
  lab} =100$ MeV. 
While the phase shift in
the $^3P_2$ channel quickly approaches the $\Lambda \to \infty$
limit,  the observed limit-cycle-like  behavior of the $^3P_0$ phase
shift reflects the non-uniqueness of solution of
Eq.~(\ref{MeqLOk0integrated}). While we still let the
possibility open to fix the solution from physical principles without
the need to rely on the data, see the discussion in Ref.~\cite{Blankleider:2001qv}, we
follow here a more pragmatic approach of Ref.~\cite{Bedaque:1998kg}.  
Specifically, we fix the solution in the $^3P_0$  partial wave by including 
a counter term of the form $C_{3P0} \,p\,p'  / \Lambda^2$ and tuning
the LEC $C_{3P0} $ to the Nijmegen partial wave analysis (PWA). 
Notice that the residual $\Lambda$-dependence of $C_{3P0}$ is of a
  logarithmic type at any finite order in the loop expansion. Consequently, it is easy to see by dimensional arguments
  that the iterations of this contact interaction do not require the
inclusion of higher-order counter terms.  Therefore, the
removed-cutoff limit
is indeed legitimate from the EFT point of view in this case, contrary
to the situation when positive powers of $\Lambda$ appear in momentum-dependent
counter terms \cite{Epelbaum:2009sd}. A more detailed analysis of
this issue will be published elsewhere. 
 
\begin{figure}
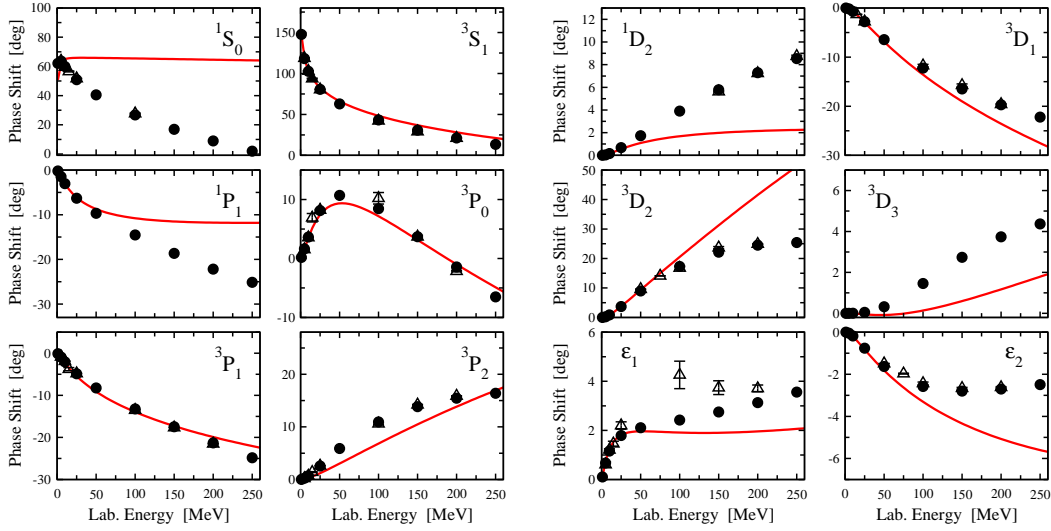

\epsfig{file=S_P.eps,width=0.48\textwidth}
\hfill
\epsfig{file=D_eps.eps,width=0.48\textwidth}
\caption[]{\label{fig:phases} Phase shifts calculated at LO in the
  modified Weinberg approach as functions of laboratory energy in
  comparison with the Nijmegen \cite{Stoks:1993tb} (filled circles) and Virginia
  Tech \cite{SAID} (open triangles) partial wave analyses. Left panel: S-
  and P-waves, right panel: D-waves and the mixing angles
  $\epsilon_{1,2}$. }
\end{figure}

We are now in the position to discuss results for phase shifts. 
We employ the exact isospin symmetry as appropriate at LO and 
use the following values for the LECs entering the OPE potential
\beq
\label{LECs}
M_\pi = 138 \mbox{ MeV}, \quad \quad 
F_\pi = 92.4 \mbox{ MeV}, \quad \quad
g_A = 1.267\,. 
\eeq
The LECs $C_{S}$, $C_{T}$ and $C_{3P0}$ are fitted to Nijmegen
$^1S_0$, $^3S_1$ and $^3P_0$ phase shifts at energies $E_{\rm lab} < 25$ MeV in the limit $\Lambda
\to \infty$. The resulting, cutoff-independent predictions for 
phase shifts in $S$-, $P$- and $D$-waves and the mixing angles
$\epsilon_{1,2}$ are visualized in Fig.~\ref{fig:phases}. 
Given that the calculations are carried out at LO, the agreement with
the Nijmegen PWA is rather good. The large deviation for the $^1S_0$
phase shift is also observed in LO KSW and (nonrelativistic) Weinberg
approach and is well-known to be largely cured by the inclusion of the 
subleading contact interaction. In all other channels, the deviations
between the theory  and Nijmegen PWA are consistent with the expected
corrections from higher-order terms in the expansion of the
potential and also indicate that these corrections can be taken into
account perturbatively. 

In addition to the predicted energy dependence of the phase shifts,
the proper inclusion of the pion-exchange physics  
can be tested in theoretical predictions for the 
coefficients in the effective range expansion
\beq
p^{2l+1} \cot \delta_l (p) = - \frac{1}{a} + \frac{1}{2} r p^2 + v_2 p^4
+ v_3 p^6 + v_4 p^8 + \ldots\,,
\eeq
where $a$, $r$ and $v_i$ denote the scattering length, effective range
and shape parameters, respectively, and $l$ is the orbital angular
momentum. 
The energy dependence
of the two-particle scattering amplitude near threshold is driven
by the long-range tale of the interaction which imposes correlations between the 
coefficients in the effective range expansion \cite{Cohen:1998jr}.
These correlations are determined by the long-range interaction and may
be regarded as low-energy theorems (LETs). In tables \ref{let_1s0} and
\ref{let_3s1}, the LETs in the  KSW and Weinberg
approaches are confronted with the results of the Nijmegen PWA 
for the $^1S_0$ and $^3S_1$ partial waves, respectively.  
\begin{table}[ttt] 
\begin{center}
\begin{tabular}{c|c|c|c|c|c}
\hline \hline 
$^1S_0$ partial wave& $a$  [fm] & $r$  [fm] & $v_2$  [fm$^3$]& $v_3$  [fm$^5$]  & $v_4$  [fm$^7$]  \\   \hline
NLO KSW from Ref.~\cite{Cohen:1998jr}  & fit   & fit  & $-3.3$  & $18$ & $-108$   
\\
LO Weinberg  &  fit  & $1.50$  & $-1.9$ &$8.6(8)$  &$-37(10)$ 
\\\hline
Nijmegen PWA & $-23.7$  & $2.67$  & $-0.5$  & $4.0$ & $-20$
\\
\hline \hline
\end{tabular}
\end{center}
\bigskip
\caption{Predictions for the coefficients in the effective
  range expansion of the $^1S_0$ phase shifts (low-energy theorems)
  with perturbative and non-perturbative treatment of the OPE
  potential in comparison with the values from the Nijmegen PWA
 (extracted using the Nijm II potential \cite{Stoks:1994wp,PavonValderrama:2005ku}).} 
\label{let_1s0}
\end{table}
Since in the KSW approach the LO S-wave amplitude does not involve
effects due to OPE, 
one needs to go to at least NLO in order to test the
LETs in this framework. The analytic expressions for the S-wave shape
parameters at NLO in the KSW scheme can be found in Ref.~\cite{Cohen:1998jr}.  
Clearly, the modified version of the KSW approach discussed in section
\ref{sec3} yields the same results for $v_i$ modulo
terms of order $1/m$ and higher.   
The LETs are known to be strongly violated in the KSW approach \cite{Cohen:1998jr}, see tables
\ref{let_1s0} and \ref{let_3s1}.  
\begin{table}[ttt] 
\begin{center}
\begin{tabular}{c|c|c|c|c|c}
\hline \hline 
$^3S_1$ partial wave& $a$  [fm] & $r$  [fm] & $v_2$  [fm$^3$]& $v_3$  [fm$^5$]  & $v_4$  [fm$^7$]  \\  \hline
NLO KSW from Ref.~\cite{Cohen:1998jr} & fit   & fit  & $-0.95$  & $4.6$ & $-25$   
\\
LO Weinberg  &  fit  & $1.60$  & $-0.05$ & $0.8(1)$  & $-4(1)$
\\ \hline
Nijmegen PWA & $5.42$  & $1.75$  & $0.04$  & $0.67$ & $-4.0$
\\
\hline \hline
\end{tabular}
\end{center}
\bigskip
\caption{Predictions for the coefficients in the effective
  range expansion of the $^3S_1$ phase shifts (low-energy theorems)
  with perturbative and non-perturbative treatment of the OPE
  potential in comparison with the values from the Nijmegen PWA \cite{deSwart:1995ui}.} 
\label{let_3s1}
\end{table}
The non-perturbative treatment of the OPE potential leads to an
improved description of the LETs in the $^1S_0$ channel. It is,
however, still rather poor at LO which should not come as a surprise
given that the long-range part of the OPE potential generates only a 
small contribution to the $^1S_0$ phase shift. One may, therefore, 
expect that the LETs are strongly affected by the two-pion exchange
contributions in this partial wave. In the $^3S_1$  channel, in
contrast, the LETs are well reproduced at LO in the Weinberg approach. The 
discrepancy for $v_2$ in the $^3S_1$ channel should not be taken too
seriously given the very small value of this coefficient. 
We further emphasize that the errors quoted for $v_{3,4}$ refer to the
estimated uncertainty of our numerical extraction of these parameters
from the  phase shifts.

\section{Summary and conclusions}
\label{sec5} 

In this paper we applied the manifestly Lorentz-invariant form of the effective
Lagrangian to the problem of nucleon-nucleon scattering without
relying on the heavy-baryon expansion. The
LO contribution to the scattering amplitude in the resulting
modified Weinberg approach can be obtained by solving  
the LS-type of integral equation (\ref{MeqLOk0integrated}) with the kernel given by
the OPE potential and derivative-less contact interactions. Contrary
to its nonrelativistic counterpart, this equation is
\emph{renormalizable}, i.e.~all UV divergences generated by its
iterations can be absorbed by redefinition of the two LO contact
interactions. The explicit appearance of the nucleon mass in the
propagators, however, makes it necessary to perform additional, finite
subtractions in order to restore the proper scaling of the
renormalized contributions in accordance with the power counting. 
Such additional subtractions only affect the values of the LECs
accompanying the LO contact interactions. Consequently, the LO
equation is renormalizable and consistent in the EFT sense. 

In the case of perturbative pions, the new approach is shown to 
reproduce the well-known results of the HB KSW framework modulo terms of
a higher order in the $1/m$-expansion.  When pions are treated
non-perturbatively as suggested in the Weinberg scheme, the
formulation we propose, being renormalizable, offers the appealing possibility to
remove the UV cutoff in the way compatible with the principles of
EFT.  We have analyzed two-nucleon scattering at
LO in the modified Weinberg approach. We found that the integral
equation does not possess a unique solution in the $^3P_0$ partial
wave similarly to the Skornyakov--Ter-Martirosyan equation for 
spin-doublet nucleon-deuteron scattering. One possible way to 
fix the solution in this channel is to include the corresponding
contact interaction whose strength is 
tuned to reproduce the low-energy data \cite{Bedaque:1998kg}. 
The obtained cutoff-independent results for phase shifts at LO in the
modified Weinberg scheme are in a reasonably good agreement with the Nijmegen
PWA. The LETs for the coefficients in the effective range expansion
are shown to be fulfilled to a very good (fair) accuracy in the
$^3S_1$ ($^1S_0$) channel. 

Clearly, the LO calculations presented here should be extended to higher
orders in the chiral expansion. Given that the main benefit of
the new formulation is its renormalizability, it is natural to treat
higher-order corrections in perturbation theory. Recent studies
\cite{Valderrama:2009ei,Valderrama:2011mv,Long:2011xw}
carried out within the nonrelativistic framework seem to suggest that 
a perturbative treatment of the two-pion exchange might be
phenomenologically successful\footnote{We do, however, not see the
  rationale behind the attempts to determine the power counting for
  short-range operators based
  on UV behavior of non-renormalizable integral equations, see
  Ref.~\cite{Epelbaum:2009sd} for an extended discussion.}. 
Also the fairly small deviations
between the Nijmegen PWA and the LO results, see
Fig.~\ref{fig:phases},  seem to be consistent with 
the possibility of a perturbative treatment of higher-order corrections.  

The proposed formulation offers also further
advantages apart from its transparency with regard to 
renormalization. In particular, one may hope to benefit from 
removing the finite cutoff artifacts which are unavoidable in the
nonrelativistic framework with non-perturbative pions. Furthermore, the avoidance of the
$1/m$-expansion can be advantageous in
situations where the momentum scale $\sim \sqrt{M_\pi m}$ associated
with radiative pions must be explicitly taken into account 
(such as e.g.~pion production in NN collisions and the quark-mass
dependence of contact interactions \cite{Mondejar:2006yu}).

\section*{Acknowledgments} 

We are grateful to Hermann Krebs, Matthias Schindler and Joan Soto
for useful discussions.   This work is supported by the EU
(HadronPhysics3 project  ``Study of strongly interacting matter''), 
the European Research Council (ERC-2010-StG 259218 NuclearEFT)
the DFG (GE 2218/2-1) and  
the Georgian Shota Rustaveli National Science Foundation (grant 11/31).


\begin{thebibliography}{99}

\bibitem{Weinberg:rz}
S.~Weinberg,
Phys.\ Lett.\ B {\bf 251}, 288 (1990); 
Nucl.\ Phys.\ {\bf B363}, 3 (1991).

\bibitem{Ordonez:1995rz} 
  C.~Ordonez, L.~Ray and U.~van Kolck,
  Phys.\ Rev.\ C {\bf 53}, 2086 (1996)
  [hep-ph/9511380].

\bibitem{Epelbaum:2008ga} 
  E.~Epelbaum, H.~-W.~Hammer and U.~-G.~Mei{\ss}ner,
  Rev.\ Mod.\ Phys.\  {\bf 81}, 1773 (2009)
  [arXiv:0811.1338 [nucl-th]].

\bibitem{Machleidt:2011zz} 
  R.~Machleidt and D.~R.~Entem,
  Phys.\ Rept.\  {\bf 503}, 1 (2011)
  [arXiv:1105.2919 [nucl-th]].

\bibitem{Epelbaum:2012vx} 
  E.~Epelbaum and U.~-G.~Mei{\ss}ner,
  arXiv:1201.2136 [nucl-th].

\bibitem{Lepage:1997cs} 
  G.~P.~Lepage,
  nucl-th/9706029.

\bibitem{Savage:1998vh} 
  M.~J.~Savage,
  In *Pasadena 1998, \emph{Nuclear physics with effective field theory}, 247-267
  [nucl-th/9804034].

\bibitem{Epelbaum:2009zz} 
  E.~Epelbaum and J.~Gegelia,
  PoS CD {\bf 09}, 077 (2009).

\bibitem{MERE} 
D.~Minossi, E.~Epelbaum, A.~Nogga, M.~Pavon Valderrama, \emph{in
preparation.} 


\bibitem{Gegelia:2004pz} 
  J.~Gegelia and S.~Scherer,
  Int.\ J.\ Mod.\ Phys.\ A {\bf 21}, 1079 (2006)
  [nucl-th/0403052].

\bibitem{PavonValderrama:2005gu} 
  M.~Pavon Valderrama and E.~Ruiz Arriola,
  Phys.\ Rev.\ C {\bf 72}, 054002 (2005)
  [nucl-th/0504067].



\bibitem{Epelbaum:2009sd} 
  E.~Epelbaum and J.~Gegelia,
  Eur.\ Phys.\ J.\ A {\bf 41}, 341 (2009)
  [arXiv:0906.3822 [nucl-th]].




\bibitem{Nogga:2005hy} 
  A.~Nogga, R.~G.~E.~Timmermans and U.~van Kolck,
  Phys.\ Rev.\ C {\bf 72}, 054006 (2005)
  [nucl-th/0506005].


\bibitem{Epelbaum:2006pt} 
  E.~Epelbaum and U.~-G.~Mei{\ss}ner,
  nucl-th/0609037.

\bibitem{Long:2007vp} 
  B.~Long and U.~van Kolck,
  Annals Phys.\  {\bf 323}, 1304 (2008)
  [arXiv:0707.4325 [quant-ph]].

\bibitem{Yang:2009pn} 
  C.~-J.~Yang, C.~.Elster and D.~R.~Phillips,
  Phys.\ Rev.\ C {\bf 80}, 044002 (2009)
  [arXiv:0905.4943 [nucl-th]].

\bibitem{Birse:2010fj} 
  M.~C.~Birse,
  [arXiv:1012.4914 [nucl-th]].

\bibitem{Valderrama:2009ei} 
  M.~P.~Valderrama,
  Phys.\ Rev.\ C {\bf 83}, 024003 (2011)
  [arXiv:0912.0699 [nucl-th]].

\bibitem{Valderrama:2011mv} 
  M.~Pavon Valderrama,
  Phys.\ Rev.\ C {\bf 84}, 064002 (2011)
  [arXiv:1108.0872 [nucl-th]].

\bibitem{Long:2011xw} 
  B.~Long and C.~J.~Yang,
  Phys.\ Rev.\ C {\bf 85}, 034002 (2012)
  [arXiv:1111.3993 [nucl-th]].

\bibitem{Kaplan:1998tg}
D.~B.~Kaplan, M.~J.~Savage, and M.~B.~Wise,
Phys.\ Lett.\ B {\bf 424}, 390 (1998).

\bibitem{Gegelia:1998ee} 
  J.~Gegelia,
  nucl-th/9806028.

\bibitem{Cohen:1998jr} 
  T.~D.~Cohen and J.~M.~Hansen,
  Phys.\ Rev.\ C {\bf 59}, 13 (1999)
  [nucl-th/9808038].

\bibitem{Fleming:1999ee} 
  S.~Fleming, T.~Mehen and I.~W.~Stewart,
  Nucl.\ Phys.\ A {\bf 677}, 313 (2000)
  [nucl-th/9911001].

\bibitem{Beane:2008bt} 
  S.~R.~Beane, D.~B.~Kaplan and A.~Vuorinen,
  Phys.\ Rev.\ C {\bf 80}, 011001 (2009)
  [arXiv:0812.3938 [nucl-th]].

\bibitem{Bernard:2007zu} 
  V.~Bernard,
  Prog.\ Part.\ Nucl.\ Phys.\  {\bf 60}, 82 (2008)
  [arXiv:0706.0312 [hep-ph]].


\bibitem{Djukanovic:2006mc} 
  D.~Djukanovic, J.~Gegelia, S.~Scherer and M.~R.~Schindler,
  Few Body Syst.\  {\bf 41}, 141 (2007)
  [nucl-th/0609055].

\bibitem{Gasser:1984yg}
J.~Gasser and H.~Leutwyler,
Ann.\ Phys.\ (N.Y.) {\bf 158}, 142 (1984).

\bibitem{Gasser:1988rb}
J.~Gasser, M.~E.~Sainio, and A.~\v{S}varc,
Nucl.\ Phys.\ {\bf B307}, 779 (1988).


\bibitem{Weinberg:1966jm} 
  S.~Weinberg,
  Phys.\ Rev.\  {\bf 150}, 1313 (1966).

\bibitem{Kadyshevsky:1967rs} 
  V.~G.~Kadyshevsky,
  Nucl.\ Phys.\ B {\bf 6}, 125 (1968).


\bibitem{Woloshyn:1974wm} 
  R.~M.~Woloshyn and A.~D.~Jackson,
  Nucl.\ Phys.\ B {\bf 64}, 269 (1973).


\bibitem{Fuchs:2003qc} 
  T.~Fuchs, J.~Gegelia, G.~Japaridze and S.~Scherer,
  Phys.\ Rev.\ D {\bf 68}, 056005 (2003)
  [hep-ph/0302117].

\bibitem{Frank:1971xx} 
  W.~Frank, D.~J.~Land and R.~M.~Spector,
  Rev.\ Mod.\ Phys.\  {\bf 43}, 36 (1971).


\bibitem{skornyakov} 
  G.~V.~Skorniakov and K.~A.~Ter-Martirosian, Sov. Phys. JETP {\bf 4} (1957) 648.

\bibitem{danilov} 
G.~S.~Danilov and V.~I.~Lebedev, Sov. Phys. JETP {\bf 17} (1963) 1015.

\bibitem{Bedaque:1998kg} 
  P.~F.~Bedaque, H.~W.~Hammer and U.~van Kolck,
  Phys.\ Rev.\ Lett.\  {\bf 82}, 463 (1999)
  [nucl-th/9809025].

\bibitem{Blankleider:2001qv} 
  B.~Blankleider and J.~Gegelia,
  AIP Conf.\ Proc.\  {\bf 603}, 233 (2001)
  [nucl-th/0107043].

\bibitem{Stoks:1993tb}
V.G.J. Stoks et~al.,
Phys.\ Rev.\ C {\bf 48}, 792 (1993).

\bibitem{SAID}
SAID on-line program, R.~A.Arndt et al., http://gwdac.phys.gwu.edu.

\bibitem{Stoks:1994wp} 
  V.~G.~J.~Stoks, R.~A.~M.~Klomp, C.~P.~F.~Terheggen and J.~J.~de Swart,
  Phys.\ Rev.\ C {\bf 49}, 2950 (1994)
  [nucl-th/9406039].

\bibitem{PavonValderrama:2005ku} 
  M.~Pavon Valderrama and E.~Ruiz Arriola,
  Phys.\ Rev.\ C {\bf 72}, 044007 (2005).

\bibitem{deSwart:1995ui} 
  J.~J.~de Swart, C.~P.~F.~Terheggen and V.~G.~J.~Stoks,
  nucl-th/9509032.

\bibitem{Mondejar:2006yu} 
  J.~Mondejar and J.~Soto,
  Eur.\ Phys.\ J.\ A {\bf 32}, 77 (2007)
  [nucl-th/0612051].


\end{thebibliography}
\end{document}